\documentclass[fleqn,usenatbib]{mnras}

\usepackage[T1]{fontenc}
\usepackage{newtxtext,newtxmath}
\usepackage{ulem}
\usepackage{soul}
\DeclareRobustCommand{\VAN}[3]{#2}
\let\VANthebibliography\thebibliography
\def\thebibliography{\DeclareRobustCommand{\VAN}[3]{##3}\VANthebibliography}

\def \rxte{{\it RXTE}}
\def \inte {{\it INTEGRAL}}
\def \swift {{\it Swift}}
\def \chandra {{\it Chandra}}

\def \src{{RX J0440.9+4431}}
\def \swiftxrt{{\it Swift}-XRT}
\def \swiftbat{{\it Swift}/BAT}
\def \nustar{{\it NuSTAR}}
\def \maxigsc{{\it MAXI}/GSC}
\def \nicer{{\it NICER}}
\def \rosat{{\it ROSAT}}
\def \maxit{{\it MAXI}}

\def \fermigbm{{\it Fermi}/GBM}

\def \astrosat{{\it AstroSat}}
\def \erg{{erg cm$^{-2}$ s$^{-1}$}}

\usepackage{graphicx}	% Including figure files
\usepackage{amsmath}	% Advanced maths commands
\usepackage[flushleft]{threeparttable}

\title[\src{} during a giant outburst]{Probing spectral and timing properties of the X-ray pulsar \src{} in the giant outburst of 2022--2023}

\author[M. Mandal et al.]
{Manoj Mandal$^{1}$, %\thanks{E-mail: manojmandal@mcconline.org.in} 
Rahul Sharma$^{2}$, %\thanks{E-mail: rsharma@rri.res.in}
Sabyasachi Pal$^{1}$\thanks{E-mail: sabya.pal@gmail.com}, 
G. K. Jaisawal$^{3}$, 
Keith C. Gendreau$^{4}$, 
 Mason Ng$^{5}$,
 \newauthor{Andrea Sanna$^{6}$},
 Christian~Malacaria$^{7}$,
 Francesco Tombesi$^{4, 8,9, 10,11}$,
 E.~C. Ferrara$^{4, 8, 12}$, 
 Craig B. Markwardt$^{4}$,
  \newauthor{Michael T. Wolff$^{13}$},
  Joel B. Coley$^{12,14}$
\\
$^{1}$Midnapore City College, Kuturia, Bhadutala, West Bengal 721129, India \\
$^{2}$Raman Research Institute, C.V. Raman Avenue, Sadashivanagar, Bengaluru, Karnataka 560080, India \\
$^{3}$DTU Space, Technical University of Denmark, Elektrovej 327-328, DK-2800 Lyngby, Denmark \\
$^{4}$Astrophysics Science Division, NASA Goddard Space Flight Center, Greenbelt, MD 20771, USA \\
$^{5}$ MIT Kavli Institute for Astrophysics and Space Research, Massachusetts Institute of Technology, Cambridge, MA 02139, USA\\
$^{6}$ Dipartimento di Fisica, Universit\`a degli Studi di Cagliari, SP Monserrato-Sestu km 0.7, 09042 Monserrato, Italy\\
$^{7}$ International Space Science Institute, Hallerstrasse 6, 3012 Bern, Switzerland \label{in:ISSI}\\
$^{8}$ Department of Astronomy, University of Maryland, College Park, MD 20742, USA \\
$^{9}$Physics Department, Tor Vergata University of Rome, Via della Ricerca Scientifica 1, 00133 Rome, Italy \\
$^{10}$ INAF – Astronomical Observatory of Rome, Via Frascati 33, 00040 Monte Porzio Catone, Italy \\
$^{11}$ INFN - Rome Tor Vergata, Via della Ricerca Scientifica 1, 00133 Rome, Italy \\
$^{12}$ Center for Research and Exploration in Space Science \& Technology II (CRESST II), NASA/GSFC, Greenbelt, MD 20771, USA \\
$^{13}$ Space Science Division, U.S. Naval Research Laboratory, Washington, DC 20375, USA\\
$^{14}$ Department of Physics and Astronomy, Howard University, Washington, DC 20059, USA\\
}
\date{Accepted XXX. Received YYY; in original form ZZZ}

\pubyear{2021}
\begin{document}
\label{firstpage}
\pagerange{\pageref{firstpage}--\pageref{lastpage}}
\maketitle

\begin{abstract}
{The X-ray pulsar \src{} went through a giant outburst in 2022 and reached a record-high flux of 2.3 Crab, as observed by \swiftbat{}. We study the evolution of different spectral and timing properties of the source using \nicer{} observations. The pulse period is found to decrease from 208 s to 205 s, and the pulse profile evolves significantly with energy and luminosity. The hardness ratio and hardness intensity diagram (HID) show remarkable evolution during the outburst. The HID turns towards the diagonal branch from the horizontal branch above a transition (critical) luminosity, suggesting the presence of two accretion modes. Each \nicer{} spectrum can be described using a cutoff power law with a blackbody component and a Gaussian at 6.4 keV. At higher luminosities, an additional Gaussian at 6.67 keV is used. The observed photon index shows negative and positive correlations with X-ray flux below and above the critical luminosity, respectively. The evolution of spectral and timing parameters suggests a possible change in the emission mechanism and beaming pattern of the pulsar depending on the spectral transition to sub- and super-critical accretion regimes. Based on the critical luminosity, the magnetic field of the neutron star can be estimated in the order of 10$^{12}$ or 10$^{13}$ G, assuming different theoretical models. Moreover, the observed iron emission line evolves from a narrow to a broad feature with luminosity. Two emission lines originating from neutral and highly ionized Fe atoms were evident in the spectra around 6.4 keV and 6.67 keV (higher luminosities).}

\end{abstract}


%%%%%%%%%%%%%%%%%%%%%%%%%%%%%%%%%%%%%%%%%%%%%%%%%%

%%%%%%%%%%%%%%%%% BODY OF PAPER %%%%%%%%%%%%%%%%%%
\section{Introduction}
\label{intro}

X-ray binaries (XRBs) contain a black hole or neutron star that is gravitationally tied to a companion star. X-ray binaries are classified into high-mass and low-mass types depending on the mass of the donor stars. In low-mass X-ray binaries (LMXBs); ($M\le 1M_\odot$), accretion takes the form of the companion's Roche-lobe overflow \citep{Re11, Sh73} whereas in high-mass X-ray binaries (HMXBs); ($M\geq 5M_\odot$), the accretion is due to the stellar wind or Be-disc of the companion star. HMXBs are strong in X-rays and contain early-type (O or B) companions. The HXMBs are classified into Be/X-ray binaries (BeXBs) and super-giant X-ray binaries (SGXBs) based on luminosity class. In the case of BeXBs, the companion star may be a dwarf, giant, or sub-giant OBe star belonging to the luminosity class III-V. BeXBs are non-supergiant systems with larger orbital periods. The SGXBs contain stars of luminosity classes I and II \citep{Re11}.

The Be/ray binaries represent roughly two-thirds of the populations of HMXBs and contain a massive non-supergiant optical companion. The Be/X-ray binary pulsar \src{} was discovered by \rosat{} in a galactic plane survey \citep{Mo97}. The long-term optical/infrared study was performed from 1995--2005 \citep{Re05}. The optical companion BSD 24-491/LS V +44 17 is a variable star with a spectral type of B0.2V, and based on optical observations, the source distance was estimated to be 3.3$\pm$0.5 kpc \citep{Re05}.  
 
\src{} went through major outbursts in 2010 (April and September) and 2011 (January). A pulsation of 202.5$\pm$0.5 s was found using {\it RXTE} \citep{Re99}. \citet{Us12} studied the source using \maxit{} and \rxte{} during the first outburst of 2010. They also reported an absorption dip in the pulse profile, probably due to the obscuration of X-ray emission by the accretion stream of the neutron star.
 The phase-resolved spectroscopy suggested a much higher absorption column density near the dip than in other phases. The source broad-band spectra were fitted using a combined model of a cutoff power law, a blackbody component, and a Gaussian component at 6.4 keV  to account for the iron emission line \citep{Us12}. 

The cyclotron resonant scattering features (CRSFs) are mostly observed in HMXBs in the energy range between 10 and 100 keV. Using the cyclotron line energy, the magnetic field  of a neutron star can be estimated directly with the 12-B-12 rule via $E_{cyc} = 11.6B_{12} (1+z)^{-1}$ keV, where z is the gravitational redshift and $B_{12}$ is the magnetic field strength in the unit of 10$^{12}$ G. Cyclotron line scattering features are subject to  variations in the geometry and dynamics of the accretion flow over the magnetic poles of NS. The luminosity dependence of the cyclotron line energy is also used to probe the accretion regimes, emission mechanisms, and beaming patterns of a pulsar \citep{Pa89, Re13}. \citet{Ts12} studied different spectral and temporal properties of \src{} using \swift{}, \rxte{}, and \inte{} during the 2010 outburst. A CRSF from the \inte{} spectrum was reported at 32 keV, and the corresponding magnetic field was estimated to be $3.2\times10^{12}$ G \citep{Ts12}. During the recent outburst in 2022, \citet{Sa23} investigated the CRSF of \src{} using \inte{} and \nustar{} observations, and the signature of the cyclotron line was not found. 
 
From the timing analysis of \src{}, \citet{Ts12} reported a pulse period of 205.0$\pm0.1$ s during the 2010 September outburst, and the pulse profile was nearly sinusoidal with a single peak feature. The pulse profile did not show any dependency on energy or luminosity, and the pulsed fraction showed a negative correlation with energy \citep{Ts12}. The energy-resolved pulse profile shows a dip-like feature near phase 0.25 at high luminosity ($\sim8\times10^{36}$ erg s$^{-1}$) which was interpreted as the change in the accretion column geometry or with the change of the orientation of the neutron star relative to the observer \citep{Ts12}.  

The X-ray pulsar went through an outburst during February 2011, and \citet{Fe13} studied different timing and spectral properties using \rxte{} and \swiftxrt{} simultaneous observations. From the reanalysis of the \inte{} data during the 2010 outburst, \citet{Fe13} reported that the 30 keV absorption feature did not improve the fit statistics significantly; alternatively, they proposed a general Comptonization model {\tt BMC} \citep{Ti97} along with an exponential high energy cutoff model to fit the spectra.

After being in quiescence for over 10 years, \src{} went through a giant outburst in December 2022, which was followed up in a multi-wavelength campaign \citep{Nak22, Ma23, Sa23, Pa23, Col23} and continued for nearly four months. We report the giant outburst from \src{} during 2022--2023. During the outburst, the X-ray pulsar was followed up by different X-ray missions \nicer{}, \nustar{}, \swiftxrt{}, \chandra{}, and \astrosat{}. During the decay phase of the outburst, an observation was conducted with the Very Large Array ({\it VLA}), but the source was not detected with a (3$\sigma$) upper limit of 57 $\mu$Jy (1.46 GHz), 6 $\mu$Jy (5.94 GHz) and 4 $\mu$Jy (8.94 GHz) \citep{Kum23}. In this paper, we have studied the  evolution of different spectral and timing parameters of \src{}  using \nicer{} observations.
The paper is organized as follows: The data reduction and analysis methods are described in Section \ref{obs}. Section \ref{res} summarises the results of spectral and timing analysis. In Section \ref{dis}, we discuss the results obtained. The findings of the study are summarised in Section \ref{con}.

 \begin{figure}
\centering{
\includegraphics[width=\columnwidth]{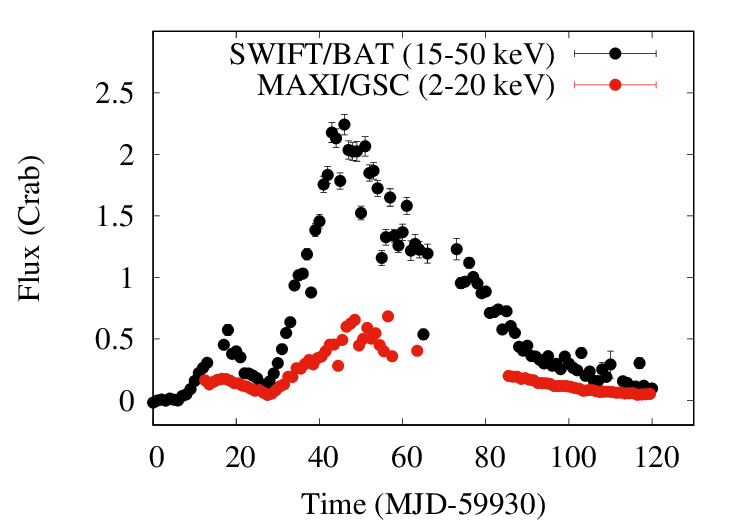}
\caption{A giant outburst is detected from \src{} using \swiftbat{} and \maxigsc{} during 2022--2023. During the peak of the outburst, the X-ray flux reached the highest value of $\sim$2.3 Crab as seen by \swiftbat{}.}
\label{fig:BAT}}
\end{figure}

\begin{figure}
\centering{
\includegraphics[width=6.25cm,angle=270]{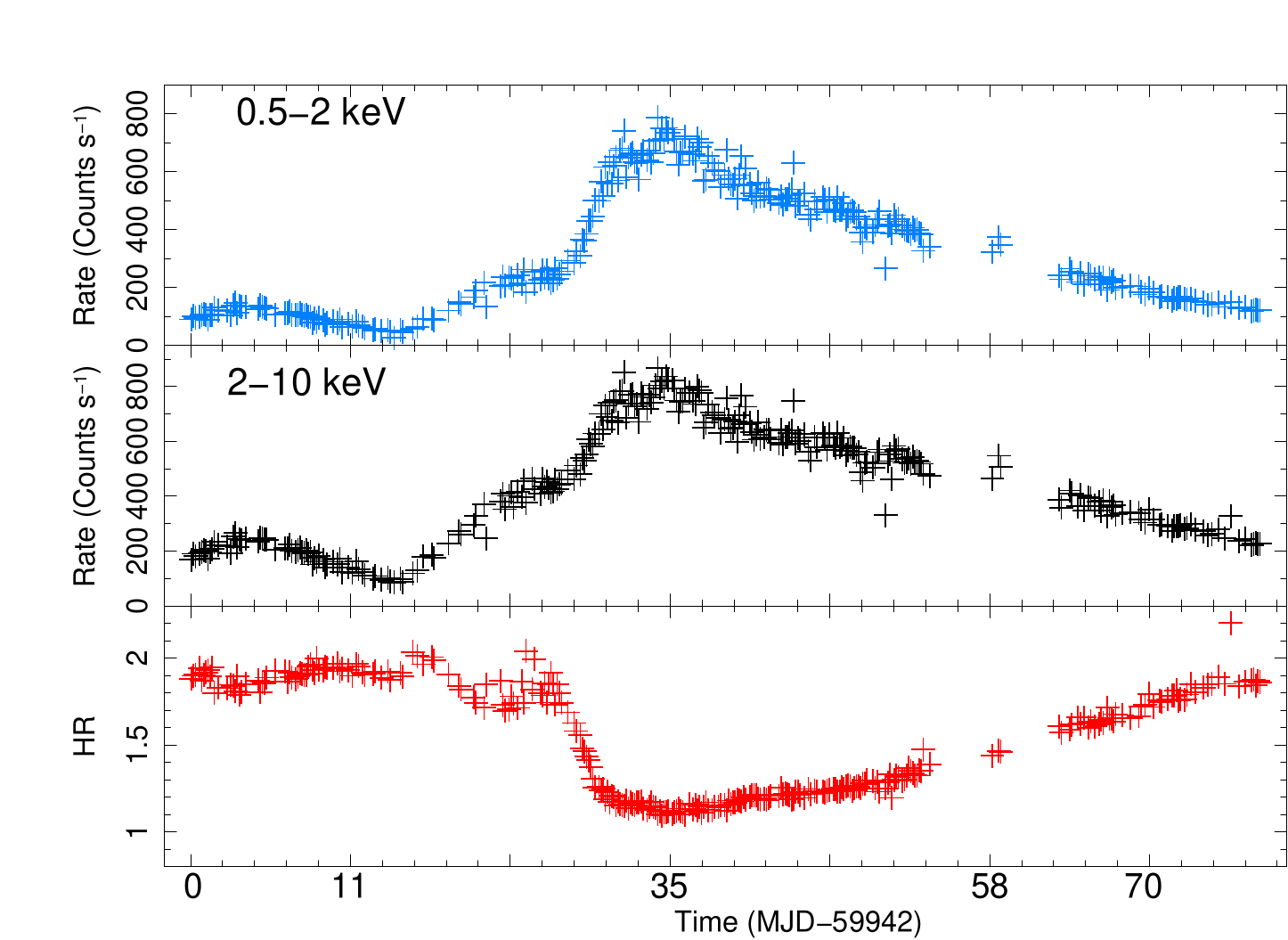}
\caption{Evolution of the hardness ratio (HR) of \src{} during the giant outburst. The first and second panels show the light curves (bin time $\sim$ 13 ks) for the \nicer{} energy bands of 0.5--2 keV and 2--10 keV, respectively, during the outburst. The bottom panel represents the evolution of the hardness ratio, which is estimated from the (2--10 keV)/(0.5--2 keV) energy band.}
	 \label{fig:HR}}
\end{figure}

\begin{figure*}
\centering{
\includegraphics[width=\columnwidth]{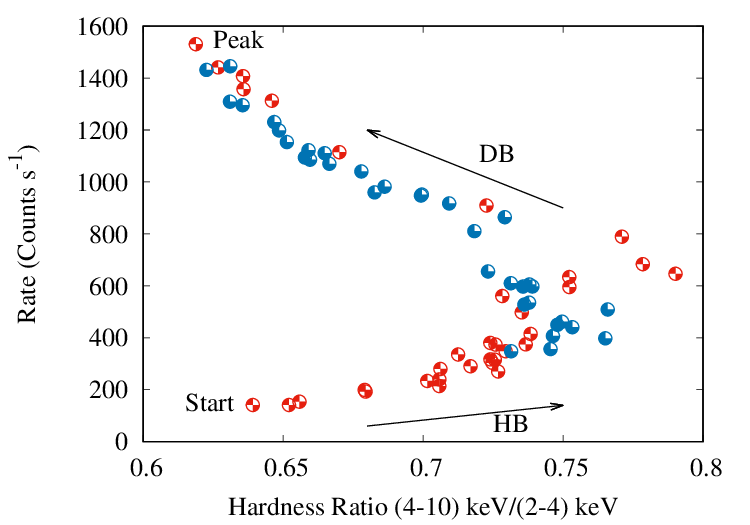}
\includegraphics[width=\columnwidth]{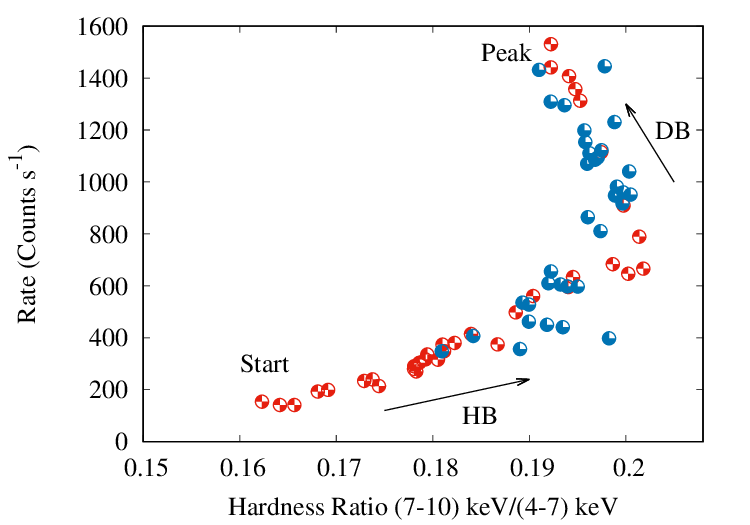}
\caption{Hardness intensity diagrams (HIDs) of \src{} during the giant outburst using \nicer{} observations for two different hardness ratios. The \nicer{} count rate is estimated for the energy range of 0.5--10 keV. The red points represent the points during the rising phase, and the blue points show the decay phase of the outburst. The HIDs show two different branches: the horizontal branch (HB) and the diagonal branch (DB). The transition from the HB to the DB took place above the critical luminosity.}
	 \label{fig:HID}}
\end{figure*}

\section{OBSERVATION AND DATA ANALYSIS}
\label{obs}
We analysed data taken by \nicer{} covering the entire outburst. The source went through an outburst (shown in Fig. \ref{fig:BAT}) from the last week of December 2022 and continued for nearly four months \citep{Nak22, Ma23, Pa23}. We reduced \nicer{} data using {\tt HEASOFT} version 6.28. We also used final data products (light curves) provided by \maxit{}\footnote{\url{http://maxi.riken.jp/star_data/J0440+445/J0440+445.html}} \citep{Ma09} and \swiftbat{}\footnote{\url{https://swift.gsfc.nasa.gov/results/transients/weak/LSVp4417/}}.

\subsection{\nicer{} observation}
The Neutron Star Interior Composition Explorer (\nicer{}) onboard the International Space Station is a non-imaging, soft X-ray telescope. The main part of {\it NICER} is the X-ray Timing Instrument (XTI), which operates in a soft X-ray region (0.2--12 keV); \citep{Ge16}. \nicer{} monitored the source from MJD 59942 and continued to observe for nearly four months during the outburst (Obs. IDs: 5203610101 to 5203610158, 6203610101 to 6203610129). The {\it NICER} data has been processed with {\tt  NICERDAS} in {\tt HEASOFT}. The cleaned event files are created by applying the standard calibration and filtering tool {\tt nicerl2} to the unfiltered data. We have extracted light curves for different energy ranges and spectra using {\tt XSELECT}. The task {\tt barycorr} is used to apply a barycentric correction for timing analysis using the JPL DE405 ephemeris and source position RA (J2000) = 70.247206766 degrees and Dec (J2000) = +44.530349458 degrees \citep{Gaia2020}.

The \nicer{} spectra are fitted in {\tt XSPEC} with the redistribution matrix file (RMF) and \nicer{} ancillary response file (ARF) provided by \nicer{} team\footnote{\url{https://heasarc.gsfc.nasa.gov/docs/nicer/proposals/nicer_tools.html}}. The good time intervals are selected for the timing analysis according to the following criteria: The ISS was not in the South Atlantic Anomaly (SAA) zone, the source elevation was $>$ 20$^\circ$ above the Earth limb, and the source direction was at least 30$^\circ$ from the bright Earth. The background corresponding to each epoch of the observation was simulated by using the {\tt nibackgen3C50}\footnote{\url{https://heasarc.gsfc.nasa.gov/docs/nicer/tools/nicer_bkg_est_tools.html}} tool \citep{Re22}.

 \begin{figure}
\centering{
\includegraphics[width=0.95\columnwidth]{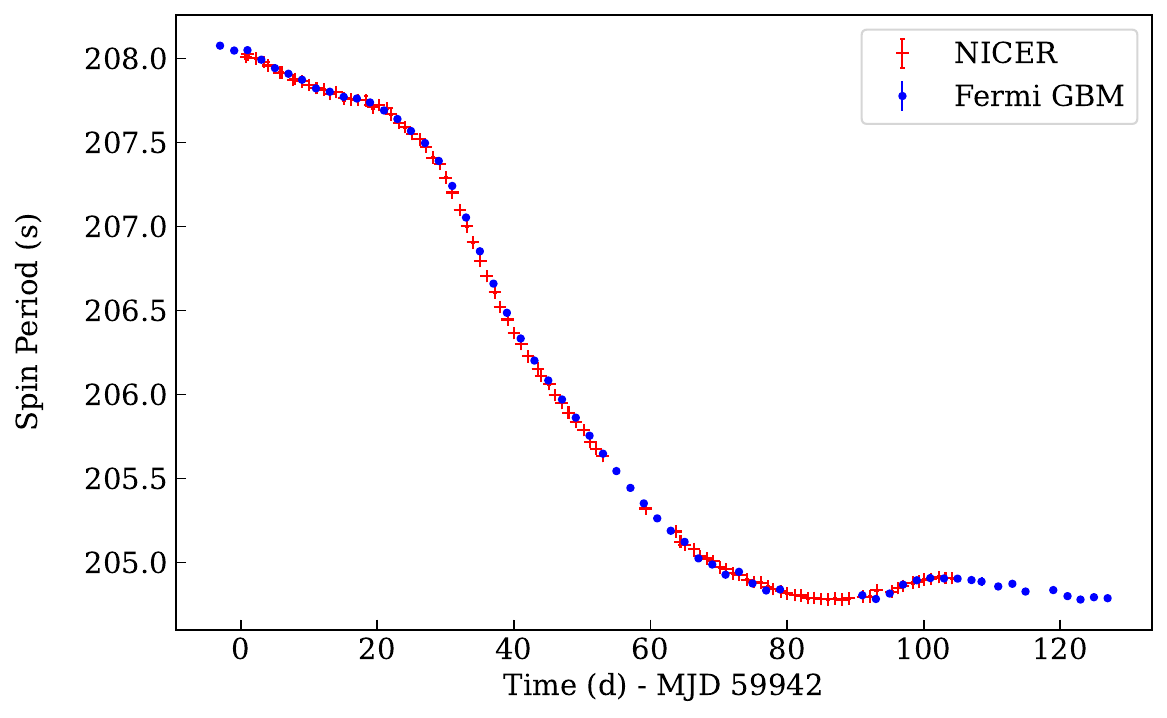}
\caption{Pulse period evolution of \src{} during the outburst using \nicer{} observations and \fermigbm{} monitoring. The error bars on spin period measurement are comparable to or smaller than the marker size.}
	 \label{fig:period}}
\end{figure}

\begin{figure*}
\centering{
\includegraphics[width=\columnwidth]{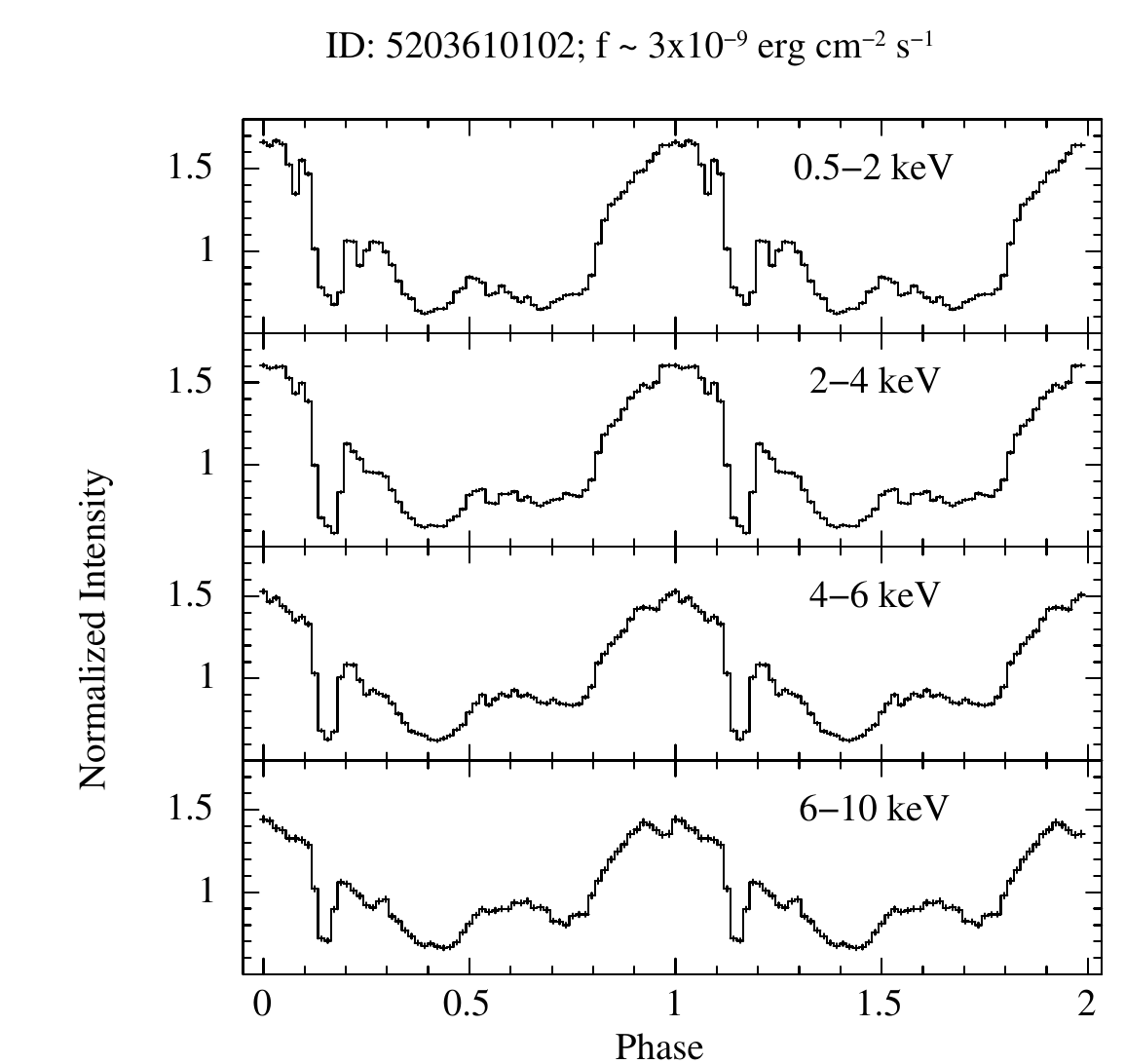}
\includegraphics[width=\columnwidth]{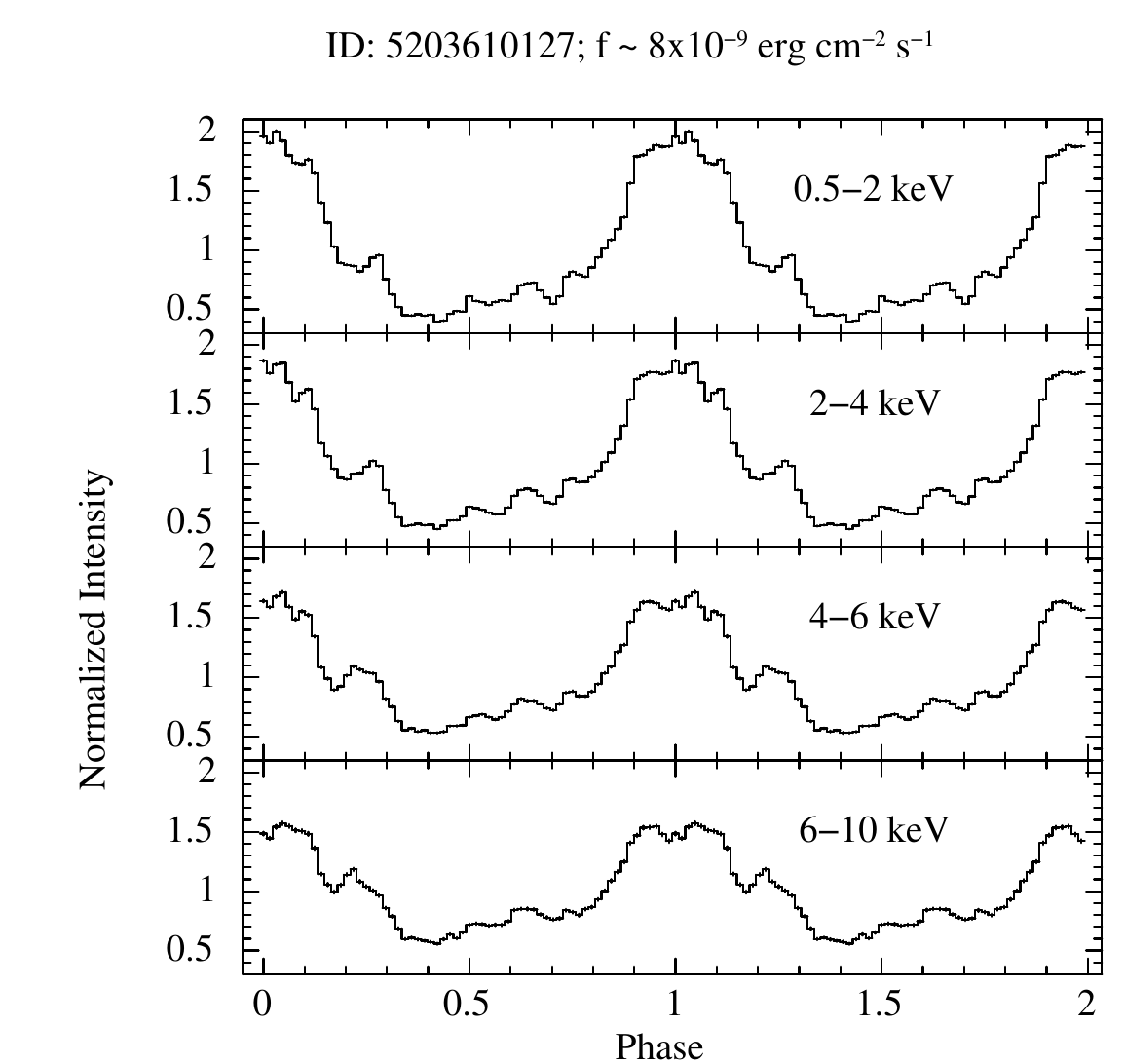}
\includegraphics[width=\columnwidth]{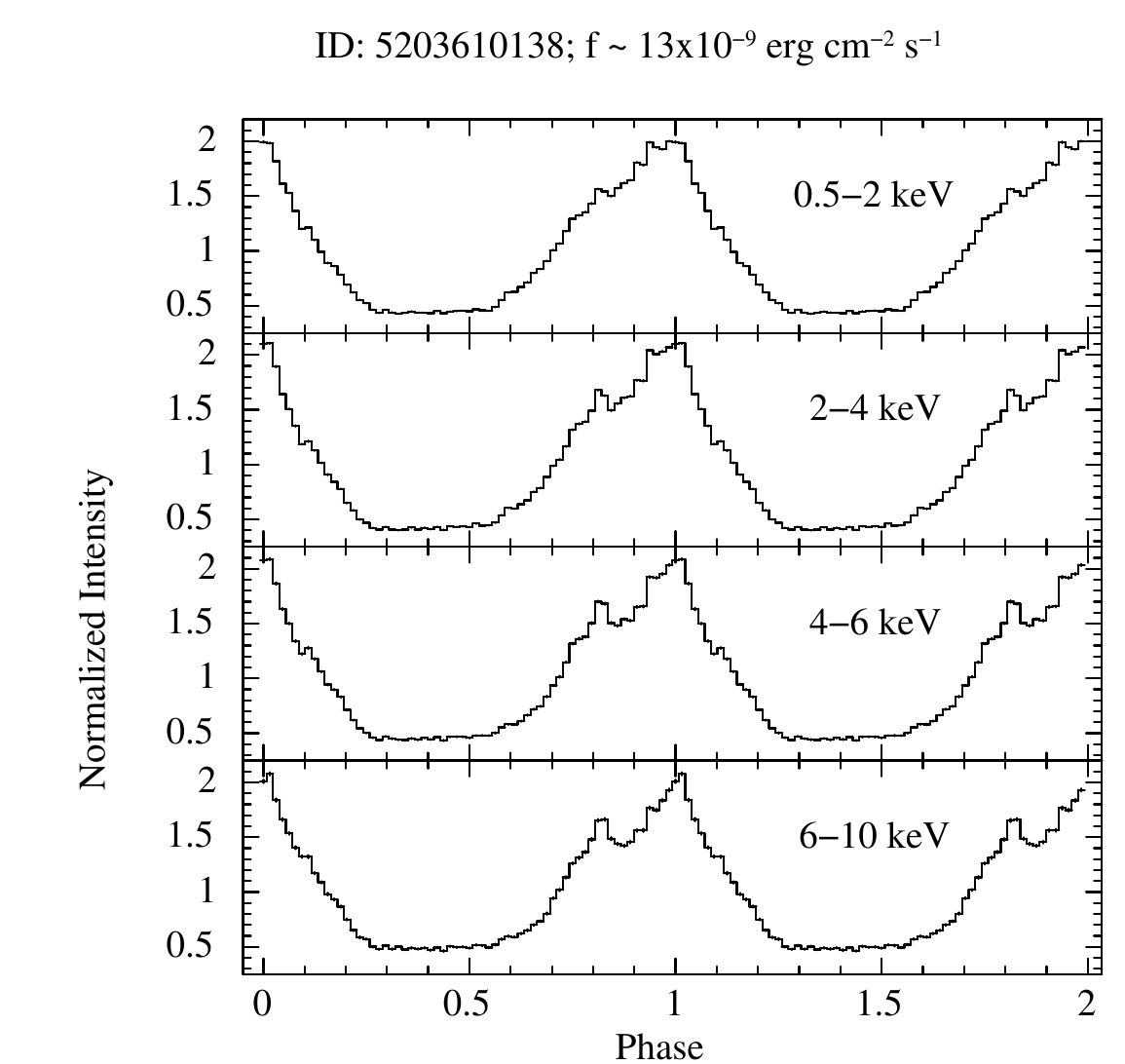}
\includegraphics[width=\columnwidth]{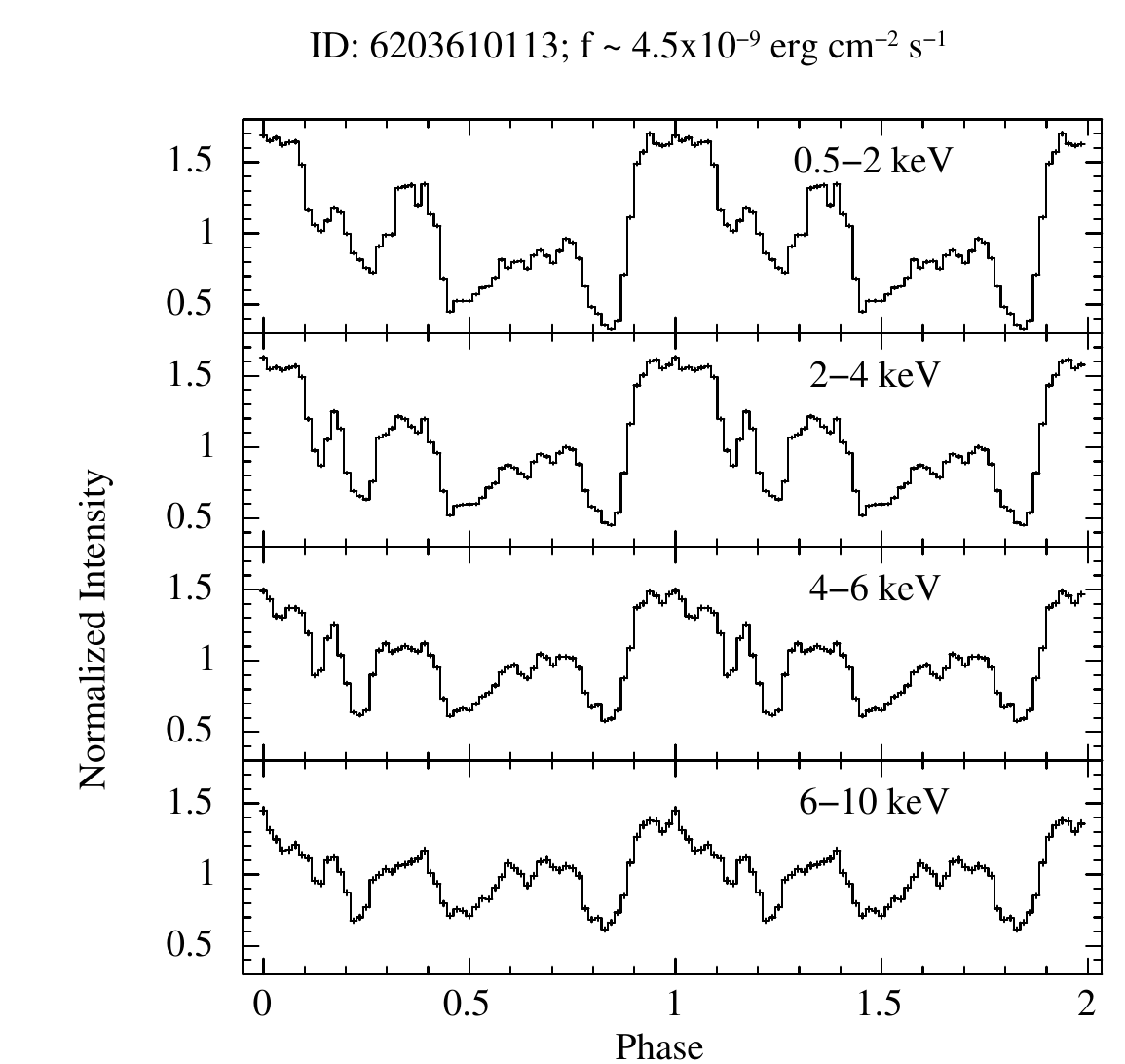}
\caption{Pulse profiles from four \nicer{} observations (IDs -- 5203610102, 5203610127, 5203610138, and 6203610113) in the 0.5--2 keV, 2--4 keV, 4--6 keV, and 6--10 keV energy ranges. These observations correspond to low flux ($\sim 3 \times 10^{-9}$ \erg), close to transition flux ($\sim 8 \times 10^{-9}$ \erg), near peak flux ($\sim 13 \times 10^{-9}$ \erg) and decay flux ($\sim 4.5 \times 10^{-9}$ \erg).}
    \label{fig:pp}}
\end{figure*}
  
\section{RESULTS}
\label{res}
In this section, we summarise the results of timing and spectral analysis of the X-ray pulsar \src{} during the 2022--2023 giant outburst. The outburst from \src{} was detected by \swiftbat{} and \maxigsc{} as shown in Fig. \ref{fig:BAT}. The peak flux reached a record high value of nearly 2.3 Crab as observed by \swiftbat{}.

\subsection{Hardness ratio evolution}
We have studied the variation in the hardness ratio of \src{} using data from different \nicer{} energy bands. The bottom panel of Fig. \ref{fig:HR} shows the variation of hardness ratio (HR), which is estimated from the ratio of count rate from two different energy bands (2--10 keV)/(0.5--2 keV). The first and second panels of Fig. \ref{fig:HR} show the variation of the source count rate in two different energy bands of 0.5--2 keV and 2--10 keV, respectively. The HR shows significant variation during the giant outburst. Initially, the HR remains at a value of $\sim$2 and during the rising phase, it shows a decreasing trend with time. At the peak of the outburst, HR attains a minimum, which corresponds to the softest state, and then HR increases again, indicating the hardening of the spectra as the outburst decays.
We also construct hardness intensity diagrams (HIDs) to investigate any state transition of the source using \nicer{} observations for different hardness ratios. Fig. \ref{fig:HID} shows the HIDs during the outburst for two different hardness ratios. The red circles represent the points during the rising phase, and the blue points show the decay part of the outburst. A transition from the horizontal branch (HB) to the diagonal branch (DB) during the supercritical regime is clearly visible in both HIDs (Fig. \ref{fig:HID}). This type of transition was observed earlier only for a few sources like KS 1947+300, EXO 2030+375, V 0332+53, 4U 0115+63, and 1A 0535+262 \citep{Re13,Ma22}. The HID suddenly turned to the left and entered the DB over the critical luminosity. The transition occurred near MJD 59969--59970 and above a luminosity of $\sim2.7\times 10^{37}$~erg\,s$^{-1}$. Hysteresis is visible in the HIDs. The hysteresis effect is more prominent in the HID, in which the hardness ratio is estimated using count rates from comparatively lower energy bands (4--10 keV)/(2--4 keV) (shown in the left-hand side of Fig. \ref{fig:HID}).

\begin{figure}
\centering{
    \includegraphics[width=\columnwidth]{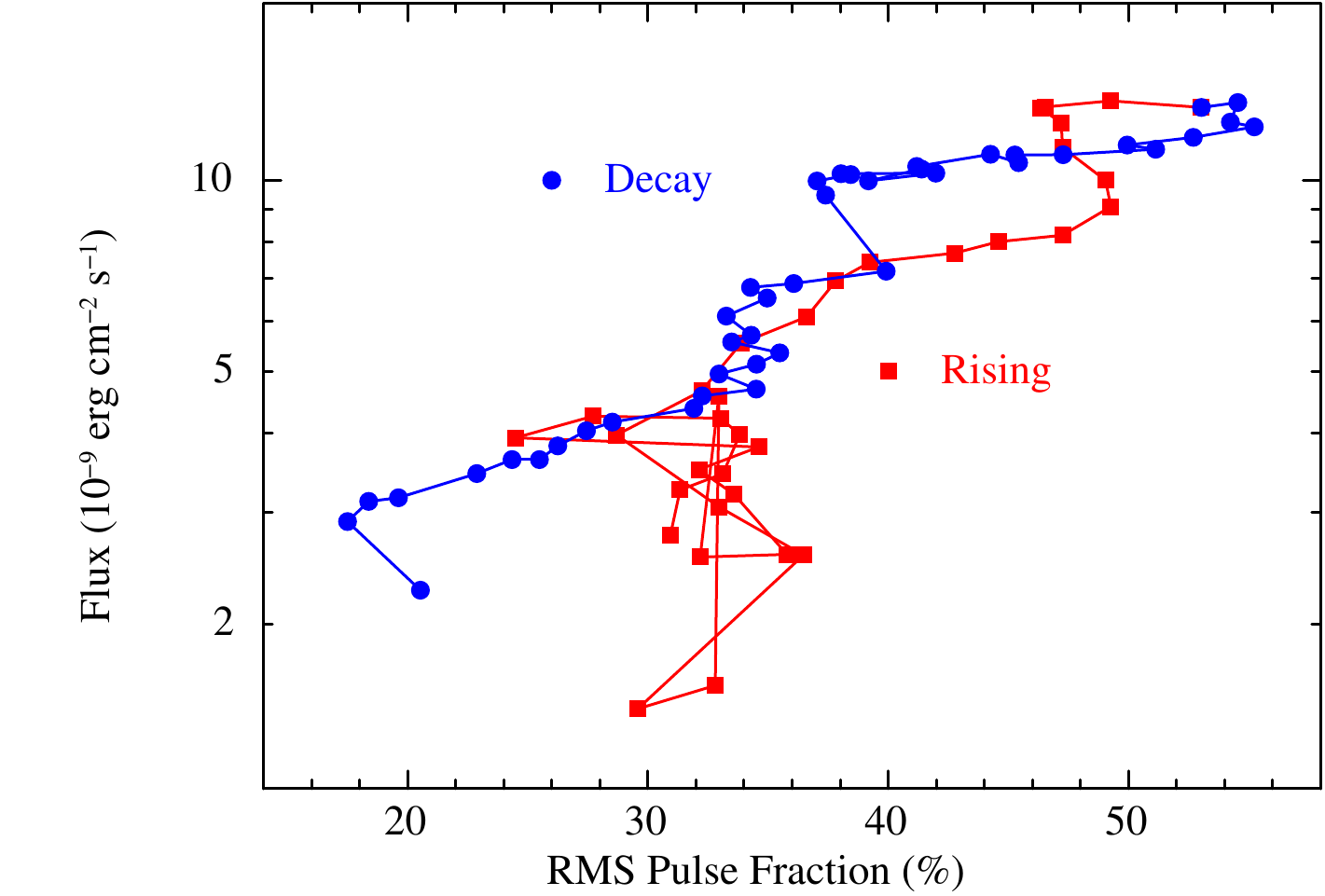}
    \caption{Variation of the rms pulsed fraction with flux during the outburst using \nicer{} (0.5--10 keV). The red points correspond to the rising phase of the outburst, and the blue points are associated with the declining phase of the outburst.}
	 \label{fig:pf}}
\end{figure}

\subsection{Evolution of pulse period and pulse profile}
Using cleaned event data, 1 s binned light curves are generated for \src{} in various energy ranges from the \nicer{} observations during the outburst. 
To search for periodicity, we used barycentre-corrected cleaned event data and performed the {\tt efsearch} task in {\tt FTOOLS}. By maximizing $\chi{^2}$ \citep{Le87} as a function of the period over 32 phase bins in each period, we folded the light curve over a trial period range to identify the best period. The uncertainties in the estimated spin periods are estimated using the bootstrap method \citep{Lutovinov12} by simulating 1000 light curves as described in \citet{Sharma23}. In the bootstrap method, we search for the best spin period on the simulated light curve in the $\sim$0.5 sec range of the best spin period obtained from the original in the step of 0.001 sec. 
Using the {\tt efold} task in {\tt FTOOLS}, pulse profiles are created by folding light curves after identifying the best spin period from observations. As the outburst progressed, \src{} showed significant evolution in the spin period. The evolution of the spin period has been found in other Be-X-ray binary pulsars over the course of the outburst due to the spin-up of neutron stars by accretion torque from accreted material. The variation of the spin period is shown in Fig. \ref{fig:period}, which indicates that the period decreases from 208 s to 205 s as the outburst evolves. 
For \src{}, the spin-up of a neutron star is clearly visible with a spin period derivative of $\sim -3.5 \times 10^{-7}$ s s$^{-1}$, but the variation is partly due to Doppler motion in a binary orbit. 

We look for the evolution of pulse profiles with energy as well as luminosity at different phases of the outburst. The source also shows a significant evolution in pulse profile shape during the outburst. The pulse profile also showed significant energy dependence. We present the pulse profiles of four \nicer{} observations at different phases of the outburst with IDs 5203610102 (MJD-59942), 5203610127 (MJD-59968), 5203610138 (MJD-59979), and 6203610113 (MJD-60017) in the energy bands of 0.5--2 keV, 2--4 keV, 4--6 keV, and 6--10 keV (Fig. \ref{fig:pp}). These profiles are created using the corresponding local spin period. The null phase for the profiles is chosen to be the phase of maximum flux. These observations correspond to low flux ($\sim 3 \times 10^{-9} $ \erg), near transition flux ($\sim 8 \times 10^{-9} $ \erg), close to peak flux ($\sim 1.3 \times 10^{-8}$ \erg) and decay flux ($\sim 4.5 \times 10^{-9}$ \erg). The pulse profiles evolve significantly with luminosity during the outburst as shown in Fig. \ref{fig:profile}. At lower luminosity, the profile shows a multi-peaked feature with an additional dip-like feature and this dip is absent in the profiles at comparatively higher luminosity. Fig. \ref{fig:pp} shows that close to the critical flux level ($\sim 8 \times 10^{-9} $ \erg), the profile evolves to a broad single peak feature. During the decay phase of the outburst, the pulse profile evolves from a single-peak feature to a multi-peaked feature.
The rms pulsed fractions evolve significantly as the outburst progresses. The rms pulsed fraction is estimated using the following formula \citep{Wilson2018}, 
\begin{equation}
    PF = \frac{1}{\Bar{p} \sqrt{N}} \left[\sum \limits_{i=1}^N (p_{i} -\Bar{p})^2\right]^\frac{1}{2}
\end{equation}
where, $p_{i}$ is count rate in the {\it i}th phase bin of pulse profile, $\Bar{p}$ is the average count rate, and $N$ is the number of phase bins.

Fig. \ref{fig:pf} shows the variation of rms pulsed fraction with flux during the outburst. The pulsed fraction increases from $\sim$18 per cent to $\sim$56 per cent as luminosity increases. The rms pulsed fractions during the rising phase of the outburst are marked in red, and the blue points represent the rms pulsed fraction during the decline phase of the outburst. The variation of the rms pulsed fraction with X-ray flux suggests a hysteresis-like pattern. Above a certain luminosity, the pulsed fraction increases with increasing X-ray flux in the rising phase of the outburst and reaches a maximum at the peak of the outburst. In the declining phase of the outburst, the pulsed fraction decreases with decreasing X-ray flux. A turnover is visible close to the critical flux level. The dependence of the pulsed fraction is less clear at lower luminosities in the rising phase. The rms pulsed fraction is also estimated for different energy ranges. The variation of rms pulsed fraction with energy at four different flux levels is shown in Fig. \ref{fig:pulse-fraction}. The rms pulsed fraction shows a trend to decrease with an increase in energy during \nicer{} observations.

\subsection{The continuum spectra}
The spectra are generated using \nicer{} observations during the giant outburst in the 1--10 keV energy band. We have used  {\tt XSPEC} \citep{Ar96} version 12.11.0 for fitting of \nicer{} spectra. The spectra are rebinned to have at least 30 counts per energy bin using  {\tt grppha}. Phenomenological power-law models with an exponential cut-off at higher energies can be used to explain the X-ray spectrum continuum \citep{Co02a}. In addition, other model components are utilized to look for any signature of emission and absorption features, such as partial covering, and Gaussian functions for emission lines from fluorescence emissions.
 
\begin{figure*}
\centering{
\includegraphics[width=5.9cm, angle=270]{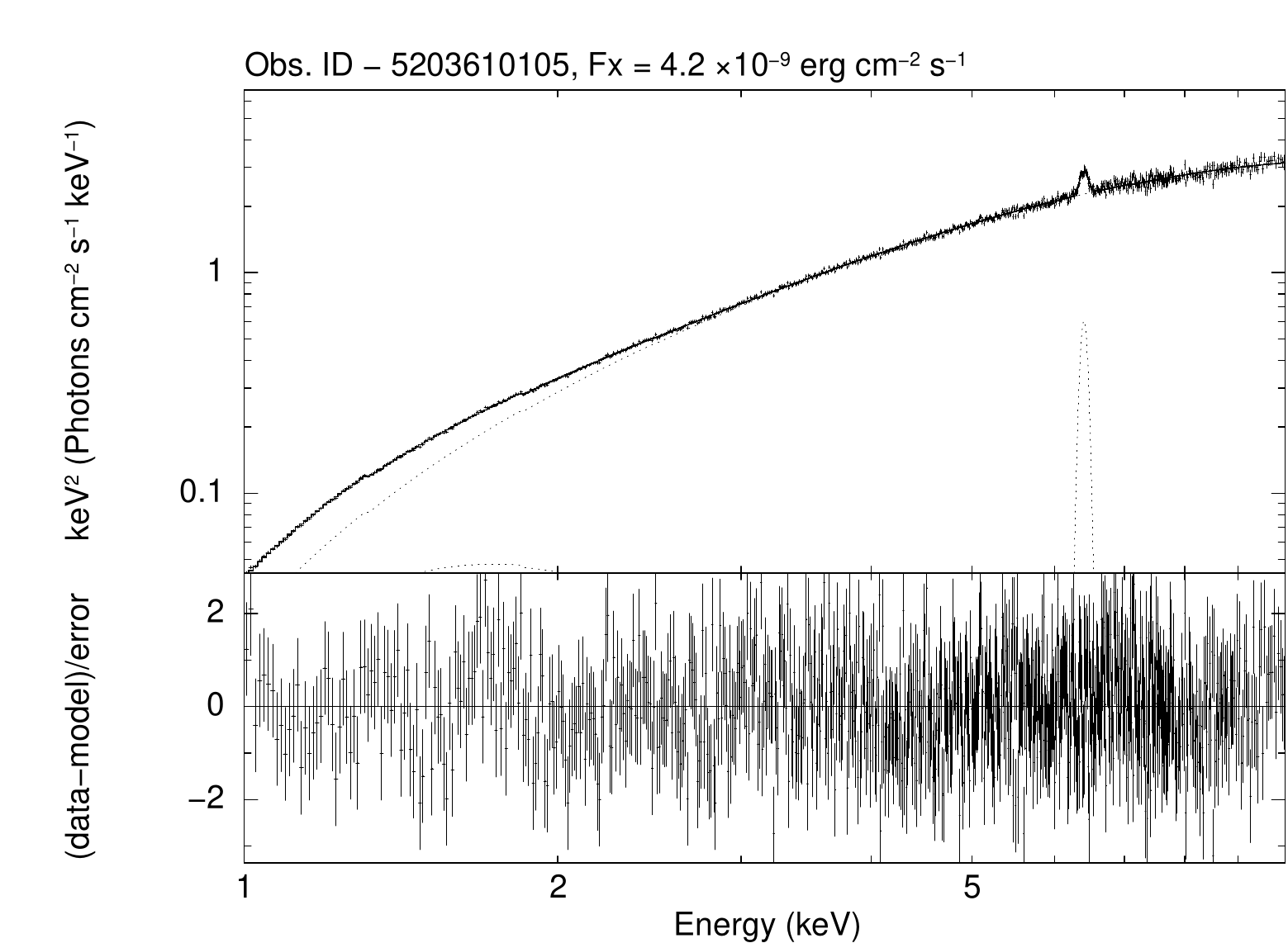} 
\includegraphics[width=5.9cm, angle=270]{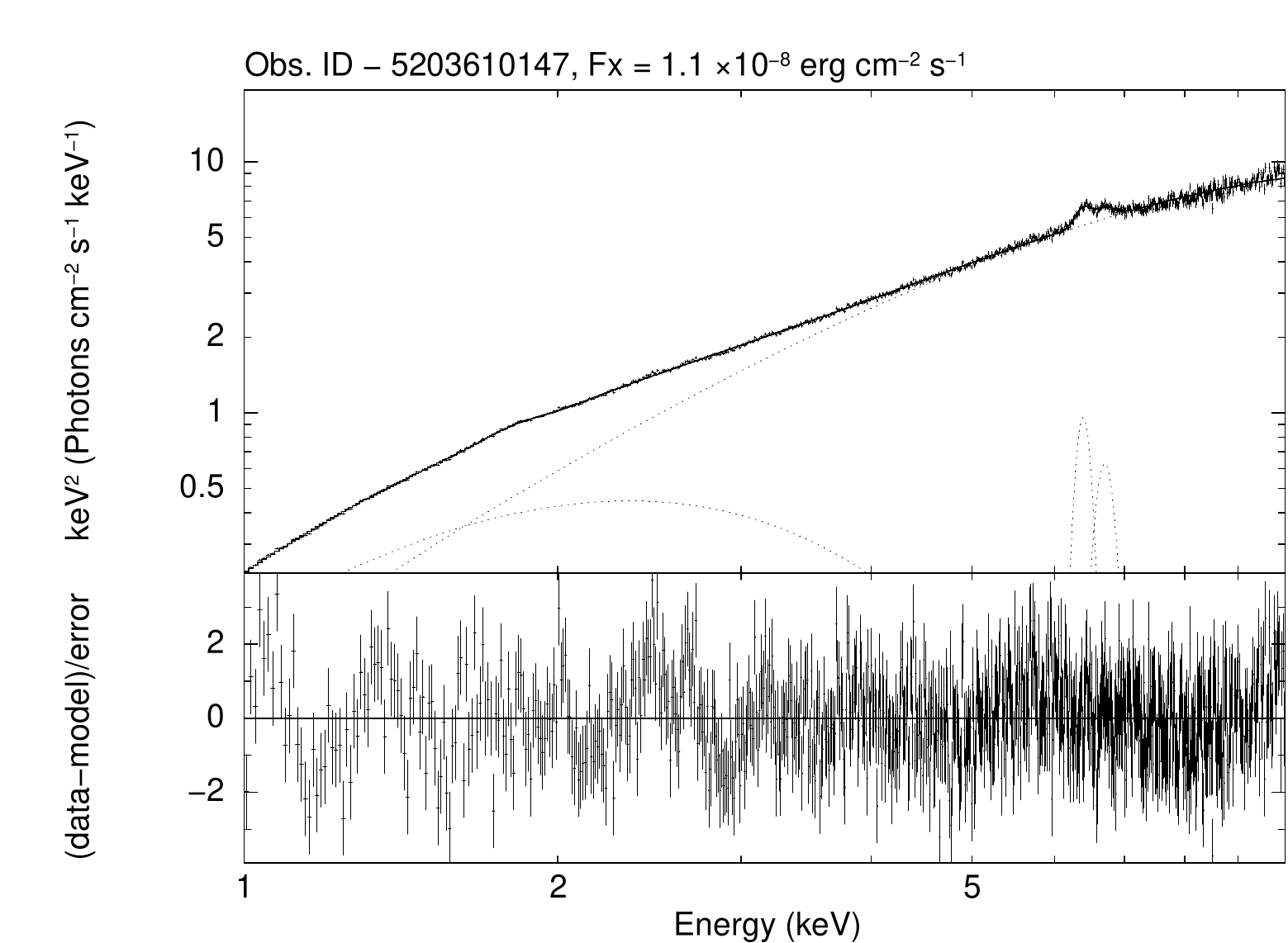} 
	\caption{Phase-averaged energy spectra of \src{} at different flux levels (indicated as $F_{x}$  in the 1--10 keV range) during \nicer{} observations. The spectra are fitted using a cutoff power-law along with a blackbody component, and an iron emission line at 6.4 keV is accounted for using a Gaussian component. At lower luminosity, a narrow iron line is fitted with a Gaussian component (as shown in the left-hand side figure), and at higher luminosity, the broad iron line feature is fitted with two Gaussian components, one at 6.4 keV with a smaller width and an additional Gaussian at 6.67 keV (as shown in the right-hand side figure).}
\label{fig:spectra}}
\end{figure*}

The \nicer{} spectra of \src{} are described by a power law with exponential cut-off ({\tt cutoffpl} in {\tt XSPEC}), which is typical for X-ray pulsar, and the presence of the soft excess in the source spectrum is modelled using an additional blackbody component {\tt bbodyrad} \citep{Us12,Ts12}.  
We have fitted the \nicer{} continuum spectra (1--10 keV) using an absorbed cutoff power law and a blackbody component. 
In addition, a Gaussian component is added to nearly 6.4 keV to account for the iron emission line. During spectral modeling at higher luminosity, there are spectral residuals in the 1.5--2.5 keV range, which is accounted for by adding a Gaussian component near 1.8 keV. This 1.8 keV feature probably originated from the instrument (Si K edge\footnote{\url{https://heasarc.gsfc.nasa.gov/docs/nicer/data_analysis/workshops/NICER-CalStatus-Markwardt-2021.pdf}}) and is not associated with the source. The best-fitting spectra at two different luminosities are shown in Fig. \ref{fig:spectra}. The left-hand side spectrum of Fig. \ref{fig:spectra} is shown for the rising phase of the outburst at low flux level ($F_X=4.2\times$10$^{-9}$ erg cm$^{-2}$ s$^{-1}$), which is modelled using absorbed cutoff power law along with a narrow iron line at 6.4 keV.  The right-hand side spectrum of Fig. \ref{fig:spectra} represents flux close to the peak of the outburst ($F_X=1.1\times$10$^{-8}$ erg cm$^{-2}$ s$^{-1}$), which  requires two Gaussian components at 6.4 keV and 6.67 keV to fit the spectrum.

The blackbody temperature varies between 0.3--0.6 keV with flux during the outburst. The spectrum is fitted using the continuum model of {\tt cutoffpl} with a photon index that varies between 0.1 and --0.9 with a cutoff energy of around 4--10 keV. 
All of the reported errors were obtained using the {\tt err} tool from {\tt XSPEC}. Uncertainties are given for a 90 per cent confidence interval. The observed peak source flux in the 1--10 keV band is $\sim1.4\times10^{-8}$ erg cm$^{-2}$ s$^{-1}$ (MJD 59977) which corresponds to the luminosity of $\sim 1 \times10^{37}$ erg s$^{-1}$ assuming a source distance of 2.44 kpc \citep{Ba21} and the corresponding bolometric peak luminosity (1--79 keV) was $\sim$ 4.2 $\times10^{37}$ erg s$^{-1}$ by using a bolometric correction of 4.15 \citep{Sa23}.

\subsection{Evolution of spectral parameters}
\begin{figure*}
\centering{
\includegraphics[width=\columnwidth]{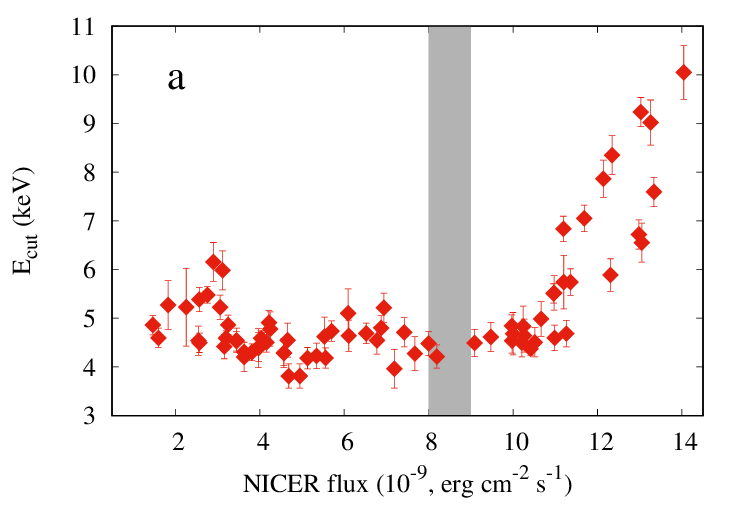} 
\includegraphics[width=\columnwidth]{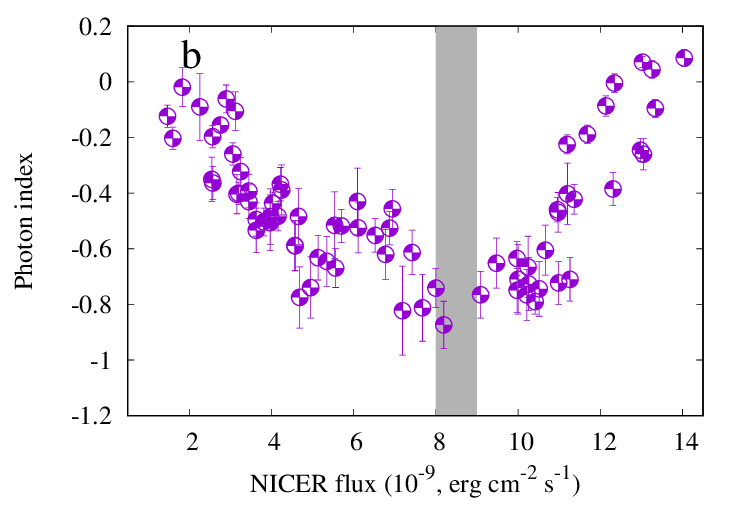} 
\includegraphics[width=\columnwidth]{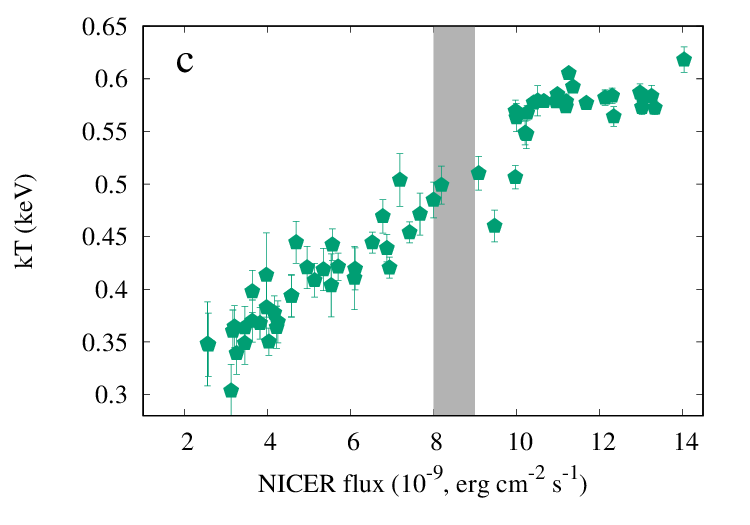} 
\includegraphics[width=\columnwidth]{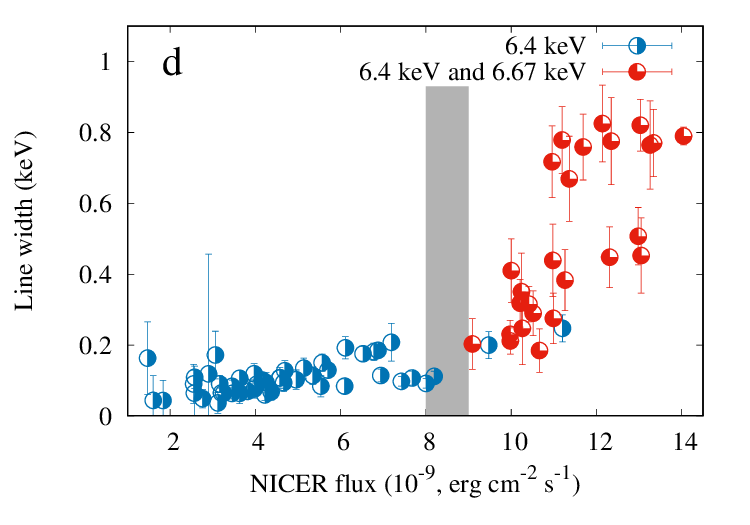} 
\includegraphics[width=\columnwidth]{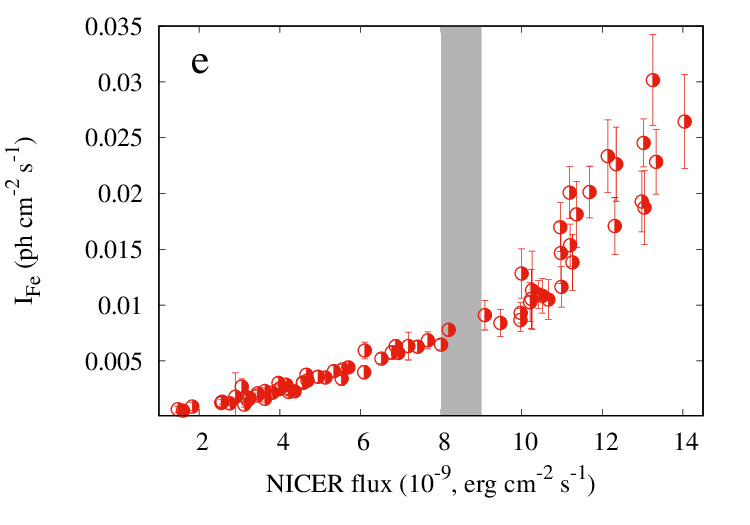}

\caption{Evolution of spectral parameters using \nicer{} observations during the outbursts. The variation of cut-off energy, photon index, kT, iron line width, and iron line flux with \nicer{} flux is shown in panels a, b, c, d, and e, respectively. The shaded region represents the critical flux level, which is also inferred from the HID. The photon index shows a negative correlation with flux below the critical flux limit, and above this limit, the correlation turns negative to positive. The correlation between the cut-off energy and flux also follows the same trend. The line width increases with flux, as seen in panel d, and the blue colour indicates the line width of only the 6.4 keV components, while the red colour represents the line width contributed by both the 6.4 keV and 6.67 keV components (total width of two Gaussian components). The iron line fluxes ($I_\textrm{Fe}$) are estimated independently by combining all Gaussian components.}
\label{fig:spectral_par}}
\end{figure*}

We look for the evolution of different spectral parameters for \src{} during the outburst using \nicer{} observations. The photon index shows a negative correlation with X-ray flux and the cutoff energy remains nearly constant below a critical flux level. The critical flux in \nicer{} 1--10 keV band lies between (8--9) $\times$ 10$^{-9}$ erg cm$^{-2}$ s$^{-1}$ which corresponds to a critical luminosity of 0.65 $\times$ 10$^{37}$ erg s$^{-1}$ assuming a source distance of 2.44 kpc and the corresponding bolometric luminosity in 1--79 keV is $\sim$2.7 $\times$ 10$^{37}$ erg s$^{-1}$. Figure \ref{fig:spectral_par} shows the variation of different spectral parameters with X-ray flux during the outburst. The shaded region represents the transition flux level, above which the photon index and cut-off energy show a change in correlation. The photon index remains close to --0.9 in the flux range of (8--9) $\times$ 10$^{-9}$ erg cm$^{-2}$ s$^{-1}$ (Fig. \ref{fig:spectral_par} (a)), which can be identified as the critical limit of the source and near this range the spectral shape changes. Above the critical flux value, the correlation between photon index and flux turns positive, and the cut-off energy also shows a similar trend (Fig. \ref{fig:spectral_par} (b)). This change in correlation between the photon index and flux above the critical flux value suggests the source state transition from the subcritical to supercritical accretion regime. 

At higher luminosities close to the peak, the hydrogen column density lies close to 0.1 $\times$ 10$^{22}$ cm$^{-2}$, which we kept frozen during the spectral fitting. The blackbody temperature shows a positive correlation with X-ray flux during the outburst. The blackbody temperature is varied in a very narrow region of 0.3--0.6 keV (Fig. \ref{fig:spectral_par} (c)), and the corresponding blackbody normalization is varied in the range of 500--1500 during the outburst. For analyzing the potential source and location of thermal emission, we have estimated the corresponding emission radius using blackbody normalization. The measured radius appears to be the maximum value of $\sim$10 km assuming a source distance of 2.44 kpc, which is about the size of a neutron star. It indicates that the thermal component could have its origins in the accretion column or the surface of the neutron star.

\begin{figure}
\centering{
\includegraphics[width=\columnwidth]{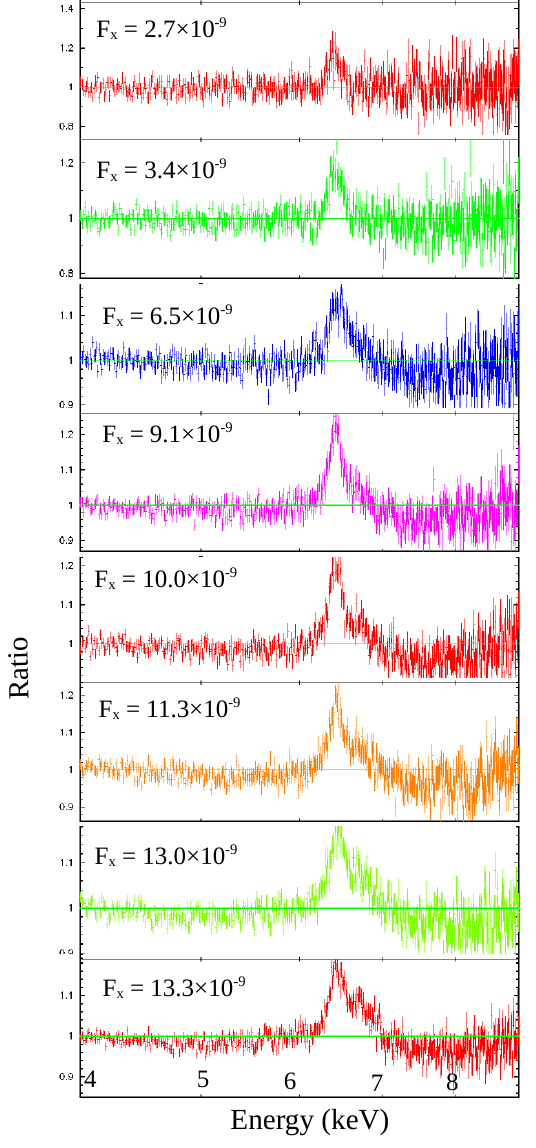} 
	\caption{Evolution of the iron line during the outburst. The ratio of \nicer{} data to the continuum model is shown. The flux ($F_X$) is estimated for the \nicer{} energy band of 1--10 keV in the units of erg cm$^{-2}$ s$^{-1}$. The figure indicates that the iron line becomes wider as the flux increases.}
\label{fig:ratio}}
\end{figure}

Strong iron emission lines are detected at 6.4 keV to 6.67 keV (above a flux level of $\ge10^{-8}$ \erg{}) using \nicer{} observations during the outburst. The iron line widths evolve significantly during the outburst. Fig. \ref{fig:spectral_par} (d) shows that the iron line width is correlated with X-ray flux, which indicates the evolution of the iron line from a narrow to a broad shape. The iron line width varies from 0.02--0.8 keV during the outburst. The evolution of the iron line feature is interesting, and a residual plot is generated to show this evolution of the feature. The iron line flux is strongly correlated with the total X-ray flux, which is shown in the bottom panel of Fig. \ref{fig:spectral_par} (e). The iron line fluxes are estimated separately by combining two individual Gaussian components at 6.4 keV and 6.67 keV. The spectral ratio between the spectral continuum model of the absorbed cutoff power law along with a blackbody component and source spectra during different phases of the outburst is shown in Fig. \ref{fig:ratio}. The iron-line feature is clearly visible in the residuals. A narrow iron line feature at $\sim$ 6.4 keV is found in lower luminosity, and at higher luminosity, this feature becomes broader and an additional Gaussian component at $\sim$6.67 keV is apparent. The iron lines are modelled using two Gaussian components at higher luminosities. The central energies of the Gaussian components are fixed to a known value of 6.4 keV and 6.67 keV, and the line width allows to vary.

\section{DISCUSSION}
\label{dis}
 We present the results from our spectral and temporal analysis using data from \nicer{} observations during the giant outburst of \src{} in 2022--2023. The giant outburst reached a record-high flux of 2.3 Crab as observed by \swiftbat{}. The source shows significant evolution in the hardness ratio, and a state transition from the subcritical to the supercritical accretion regime is detected from the HID. The HB to DB transition for the pulsar \src{} is noticed in the HID (Fig. \ref{fig:HID}). Earlier, several sources, including EXO 2030+375, 4U 0115+63, V 0332+53, 1A 0535+262, and KS 1947+300, showed similar types of patterns in the HID \citep{Re13,Ma22}. At the critical luminosity, \src{} suddenly turned (to the left) in the HID and entered the diagonal branch. The hardness ratio began to drop above the critical luminosity, and the peak of the massive outburst corresponded with the softest state of the DB. The horizontal branch and diagonal branch are two distinct branches that appear in hardness-intensity diagrams during the transition from the subcritical to the supercritical domain, respectively. The HB represents the low-luminosity state, indicated by high X-ray variability and spectral variations. When the X-ray luminosity exceeds the critical limit, the DB correspondingly represents the high-luminosity condition. Depending on the HID patterns, the classifications HB and DB are established \citep{Re13}. In the subcritical and supercritical regimes, respectively, the HB and DB patterns are typically seen. The HIDs of \src{} show hysteresis effect, although not as significant as 1A 0535+262, 4U 0115+63 \citep{Reig22, Re08}. Hysteresis primarily refers to the phenomenon where certain spectral or temporal parameters at a given luminosity level exhibit different values based on whether the source is in the rising or declining phase of the outburst. Hysteresis in HIDs refers to different values of spectral hardness at the same luminosity level depending on whether the source is in the decay or rising phase of the outburst. The hysteresis in HIDs was observed earlier for several sources like 4U 0115+63, V0332+53, EXO 2030+375, and 1A 0535+262, etc. \citep{Re08, Reig22}. Earlier, it was found that the hysteresis is most prominent in the HID using low energy bands, and the patterns disappear when comparatively higher energy bands (above 4 keV) are considered to estimate the hardness ratio \citep{Reig22}. The hysteresis patterns were also seen in the spin rate \citep{Fil17}, and pulsed fraction \citep{Wa20} for different sources.
 
The evolution of pulse period, pulse profile, and pulsed fraction is studied during the outburst using \nicer{} observations. The spin period decreases slowly as the burst evolves. The pulse profiles are found to be highly variable with luminosity and energy. The evolution of the pulse profile from a multi-peaked feature to a broad single peak near the critical flux level is observed. This change in pulse profile may indicate a change in beaming patterns, which is associated with the state transition of the source. At lower luminosities, the pulse profiles show a multi-peaked feature with an additional absorption dip, which is most probably due to the obscuration of X-ray emission by the accretion stream of the neutron star. A similar feature was also reported in an earlier outburst using \rxte{} by \citet{Us12}. At the peak of the outburst, the pulse profile shows a broad single peak feature with an additional energy-dependent hump. During the decay phase, the pulse profile again shows a multi-peaked feature with strong energy dependence. The dependence of the pulsed fraction on luminosity is sometimes used to probe the accretion regime. However, the dependence of the pulsed fraction on luminosity is not unique. In some sources, the pulsed fraction exhibited a negative correlation with luminosity in the subcritical accretion domain \citep[e.g., Swift J0243.6+6124, V 0332+53, EXO 2030+375;][]{Wilson2018, Lu09} whereas, in some cases, the correlation was opposite \citep[e.g., 1A 0535+262;][]{Ma22}. In our observation, in the subcritical domain, the pulsed fraction was positively correlated with X-ray flux at low luminosity. Near the critical luminosity, a sudden turn is visible (in the rising phase). Below the critical luminosity, the pulsed fraction is larger in the rising phase compared to the decay phase. But, above the critical luminosity, the pulsed fraction becomes smaller in the rising phase compared to the decay phase. Such a phenomenon was also observed earlier for Swift J0243.6+6124, which may indicate a transition between accretion modes \citep{Wa20}. The mechanism related to the luminosity dependence of the pulsed fraction is still unclear. To understand the association between pulsed fraction and luminosity for various sources, a more detailed model needs to be taken into account.

The luminosity of a pulsar affects the beaming patterns. To distinguish two different accretion regimes of the source, the critical luminosity is crucial. In the subcritical domain, the source luminosity is lower than the critical luminosity ($L_x=10^{34-35}$ erg s$^{-1}$). In the subcritical regime, accreting material falls freely on the surface of the neutron star, which leads to a `pencil beam' X-ray emission \citep{Ba75}. Emission escapes from the top of the column in the pencil beam pattern \citep{Bu91} and in this scenario, the photons propagate along the magnetic field lines. At higher luminosities ($L_x\ge10^{37}$ erg s$^{-1}$), accretion-dominated shock is supposed to be formed near the critical luminosity, which results in a spectral state transition and change in beaming patterns. In the supercritical regime, the radiation pressure is high enough to halt accreting matter at a certain distance above the neutron star and produce a radiation-dominated shock \citep{Ba76, Be12}. In the higher luminosity state, the beaming patterns are mostly dominated by fan-beam patterns or a mixture of pencil and fan-beam patterns.

The \nicer{} spectral fitting results suggest that the spectral parameters show a significant evolution during the outburst. A spectral transition for \src{} is observed during the outburst. The correlation between photon index and \nicer{} flux turns from a negative to positive trend above the flux range of (8--9) $\times$ 10$^{-9}$ erg cm$^{-2}$ s$^{-1}$ in the 1--10 keV range. Above the corresponding critical luminosity of $\sim$2.7 $\times$ 10$^{37}$ erg s$^{-1}$, the emission mechanism and beaming pattern are supposed to be changed. The shock plays a dominant role in the accretion column, effectively reducing the velocity of energetic electrons, which accounts for the positive correlation between the photon index and luminosity in the supercritical regime. As a result, the source spectrum becomes softer due to the lack of bulk Comptonization of photons with accreting electrons \citep{Be12}.
The magnetic field can be estimated using the critical luminosity. We suggest the luminosity where the spectral transition takes place for \src{} is associated with the critical luminosity. For a typical neutron star (mass of 1.4 $M_\odot$,  radius 10 km), the magnetic field is related to the critical luminosity by \citet{Be12}; assuming disc accretion onto a classical neutron star ($\Lambda$ is a constant, depends on the accretion flow configuration, and taken as $\Lambda$ = 0.1),
%
%e8 #&#
\begin{equation}
L_{\mathrm{critical}} = 1.5 \times 10^{37}\left (\frac{B}{10^{12} G}
\right )^{\frac{16}{15}}~\text{erg}\,\text{s}^{-1}
\end{equation}
The magnetic field corresponding to the critical luminosity $2.7\times 10^{37}$~erg\,s$^{-1}$ is estimated to be in the order of $10^{12}$~G for a source distance of 2.44 kpc. 

 The critical luminosity depends on various parameters like the radius, mass, and accretion flow geometry of the neutron star \citep{Mu15b}. The geometry of the accretion channel and the radius of the magnetosphere also have an impact on the critical luminosity. We have also used the model for critical luminosity and magnetic field as suggested by \citet{Mu15b} for different scenarios to get an alternative estimation of the magnetic field. 
 The impact of polarization is strong at higher magnetic fields as the cross-section varies with the mode of polarization (X or O mode). In a high magnetic field (above $10^{12}$ G), the scattering cross section for different photon polarizations is very different (see Fig. 5e of \citet{Mu15b}). To estimate the magnetic field, we used calculations by \citet{Mu15b} which consider the  polarization, resonance in the Compton scattering cross-section, and the configuration of the accretion flow. Assuming the mass accretion through the disc ($\Lambda< 1$, taken as 0.5 here), electron temperature ($T_e$) = 1 keV, $\xi$-coefficient = 1.5, $l_0$/l = 0.5, $M$ = 1.4 $M_\odot$, $R$ = 10$^6$~cm, and for a pure X-mode of polarization, the expected cyclotron line should be above 100 keV (for a critical luminosity of $2.7\times 10^{37}$~erg\,s$^{-1}$) and the corresponding magnetic field is in the order of 10$^{13}$ G. This provides an order of magnitude higher for magnetic field strength in comparison to the estimation made using \citet{Be12}.

In the subcritical regime, the negative correlation between photon index and X-ray flux implies the hardening of the power-law continuum with flux. In this scenario, the X-ray emission may originate from the hot mound of the neutron star's surface \citep{Ba76, Be12}. Earlier, several sources showed a significant variation in the L~--$~\Gamma$ diagram close to critical luminosity. The transition from a negative to positive correlation was seen in the L~--$~\Gamma$ diagram as luminosity increased \citep{Re13}.
Earlier, in the subcritical regime, a negative correlation was reported for the sources like 1A 1118--612, GRO J1008--57, XTE J0658--073, and a change in the correlation of $L$~--$~\Gamma $ was found for the sources 1A 0535+262 \citep {Ma22}, and KS 1947+300 \citep{Re13}, 4U 0115+63, SMC X-2 \citep{Jai23}, EXO 2030+375 \citep{Ep17, Ja21}, and 2S 1417--624 \citep{Se22}. The change in correlation in the L~--$~\Gamma$ diagram suggests a spectral transition from the sub-critical to the super-critical regime. The cutoff energy also shows a significant evolution during the state transition of \src{}. The cutoff energy nearly follows the same trend as the photon index. Above the critical luminosity, the cutoff energy increases with flux, and at the peak of the outburst, the cutoff energy reaches its highest value. The higher values of the cutoff energy correspond to the softer spectra of the source. The evolution of cutoff energy with luminosity is used to probe the accretion regime of a source in an outburst. Earlier, several sources showed a change in correlation between the cutoff energy and luminosity. In the diagonal branch, a positive correlation between the cutoff energy and flux was reported for KS 1947+300, V 03332+53, EXO 2030+375, GX 304-1, SMC X-2, and 1A 0535+262 \citep{Jai23,Ep17,Re13,Ma15,Ma22}.

We report iron emission lines and the evolution of the iron lines with luminosity for \src{}. A strong iron line is detected from \nicer{} observation and the line width increases with the increase in flux. The 6.4 keV line energy does not vary significantly during the outburst, which may be associated with near-neutral material that may be the composite lines between Fe II and Fe XVIII close to 6.4 keV \citep{Re13}. The Fe K$\alpha$ line may originate in relatively cool matter from the reprocessing of the hard X-ray continuum. The 6.4 keV line forms near the neutral iron, and with the increase in ionization, the line energy increases, resulting in 6.67 keV and 6.98 keV lines. Due to a small separation in energy, even the ionized iron up to XVIII may be a part of the 6.4 keV blended line \citep{Eb96, Li05}. The narrow iron line of \src{} evolved to a broad shape feature with the increase in luminosity. An additional emission component arose at 6.67 keV along with the 6.4 keV component at higher luminosity.
Earlier, for the ultra-luminous X-ray pulsar Swift J0243.6+6124, the iron line profile evolved significantly with luminosity, and the broadening of the line profile was contributed by the 6.67 keV (Fe XXV) and 6.97 keV (Fe XXVI) \citep{Ja19}. The 6.67 keV lines may appear from highly ionized He-like Fe ions, and the 6.4 keV  fluorescent line may appear from neutral or weakly ionized Fe. The evolution of iron line flux with the total X-ray flux is investigated during the outburst. The iron line flux strongly correlates with X-ray flux, which implies that the increase in line emission and the line strength are probably due to the illumination of the cool matter. Earlier, a strong correlation was found between the iron line flux and X-ray continuum flux for several sources like EXO 2030+375, V 0332+53, KS 1947+300, 1A 0535+262, GRO J1008-57, 4U 0115+63, 1A 1118-612, Swift J1626.6-5156, and XTE J0658-073 \citep{Re13}. The strong correlation of iron line flux with X-ray continuum flux indicates the contribution of thermal hot plasma present along the Galactic plane is not significant \citep{Ya09}.

\section{CONCLUSIONS}
\label{con}
We report the giant outburst from \src{} with a record high flux of 2.3 Crab as observed by \swiftbat{}. The detailed spectral and timing properties are explored using \nicer{} data. The pulse profiles are found to be highly luminous and energy-dependent. The evolution of the pulse profile with luminosity at different phases of the outburst indicates a change in the beaming patterns and emission mechanism of the source. The hardness ratio and hardness intensity diagram indicate a state transition from subcritical to supercritical accretion regimes. A transition from a horizontal to a diagonal branch is detected during the giant outburst. The state transition is also confirmed by the evolution of spectral parameters. The photon index shows a transition from negative to positive above the critical luminosity. The critical luminosity is found to be $\sim$2.7 $\times$ 10$^{37}$ erg s$^{-1}$, above which the state transition is triggered. The magnetic field of the neutron star can be estimated in the order of 10$^{12}$ or 10$^{13}$ G assuming different theoretical models. The source spectra can be modelled using an absorbed cutoff power law along with a blackbody component, a Gaussian at 6.4 keV, and an additional Gaussian at 6.67 keV used at higher luminosity. The strong iron emission line evolved from a narrow to a broad feature with the increase in luminosity. The iron line flux is found to be strongly correlated with the X-ray flux. Two emission lines from neutral and highly ionized Fe atoms at 6.4 and 6.67 keV are present in the Fe band. 

\section*{Acknowledgements}
We thank the anonymous reviewer for their useful suggestions, which helped improve the manuscript significantly. This research has made use of the \maxit{} data provided by RIKEN, JAXA, and the \maxit{} team. We are also thankful to the \nicer{} team for efficient continuous monitoring of the source. ECF and J.B. Coley are supported by NASA under award number 80GSFC21M0002. Astrophysics research at the Naval Research Laboratory is supported by the NASA Astrophysics Explorer Program.

%%%%%%%%%%%%%%%%%%%%%%%%%%%%%%%%%%%%%%%%%%%%%%%%%%
\section*{Data Availability}
The data underlying this article are publicly available in the High Energy Astrophysics Science Archive Research Center (HEASARC) at \url{https://heasarc.gsfc.nasa.gov/db-perl/W3Browse/w3browse.pl}. 


\bibliographystyle{mnras}
%\bibliography
{example} % if your bibtex file is called example.bib

%%%%%%%%%%%%%%%%%%%%%%%%%%%%%%%%%%%%%%%%%%%%%%%%%%

%%%%%%%%%%%%%%%%% APPENDICES %%%%%%%%%%%%%%%%%%%%%

\appendix

\section{Pulse profile and pulsed fraction}
\begin{figure}
\centering{
\includegraphics[width=0.9\columnwidth]{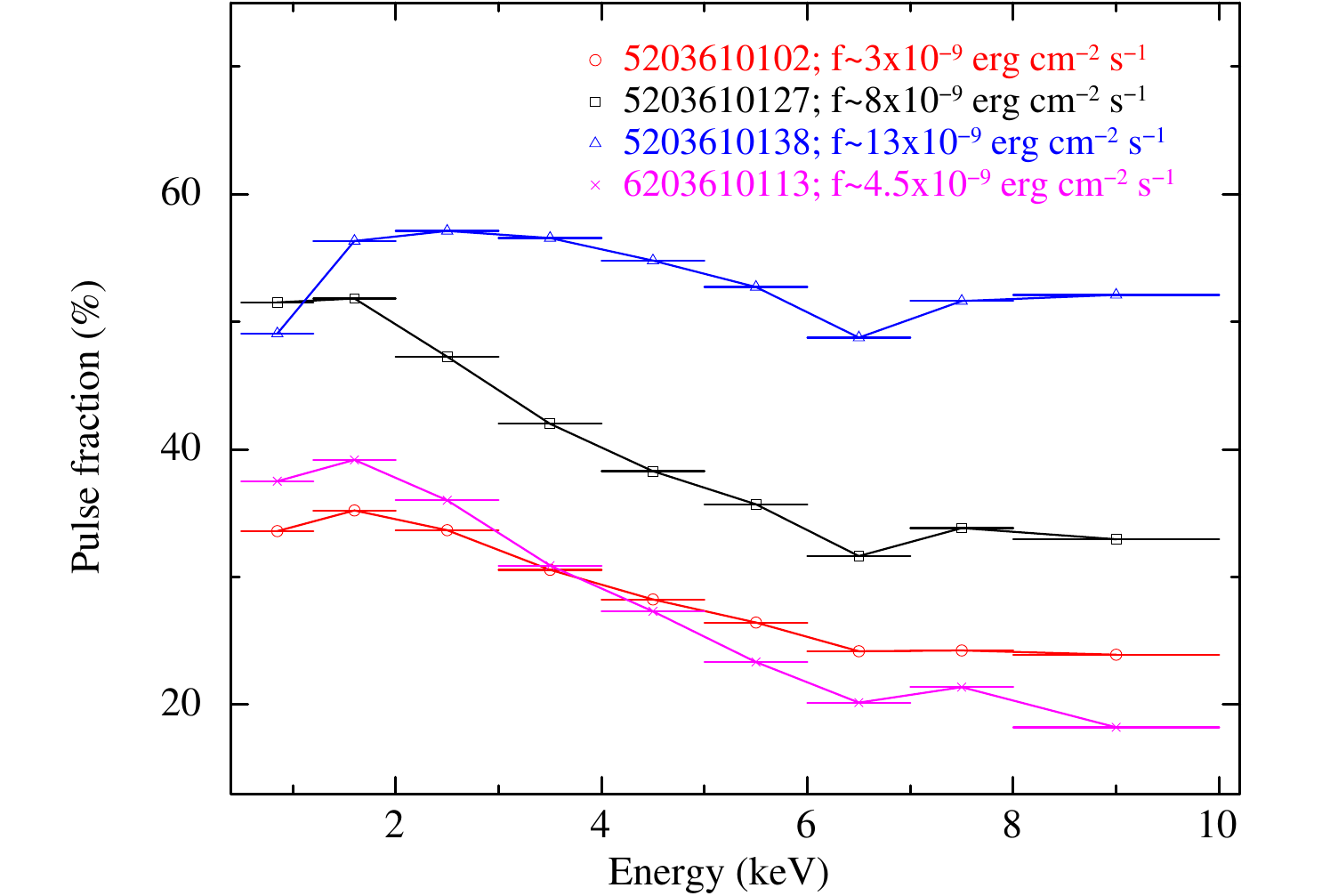}
\caption{Variation of rms pulsed fraction with energy for different flux levels. Pulse profiles from four \nicer{} observations 5203610102, 5203610127, 5203610138, and 6203610113 correspond to low flux ($\sim 3 \times 10^{-9}$ \erg), close to transition flux ($\sim 8 \times 10^{-9}$ \erg), near peak flux ($\sim 1.3 \times 10^{-8}$ \erg) and decay flux ($\sim 4.5 \times 10^{-9}$ \erg) are created in different energy ranges.}
	 \label{fig:pulse-fraction}}
\end{figure}

\begin{figure}
\centering{
\includegraphics[width=0.9\columnwidth]{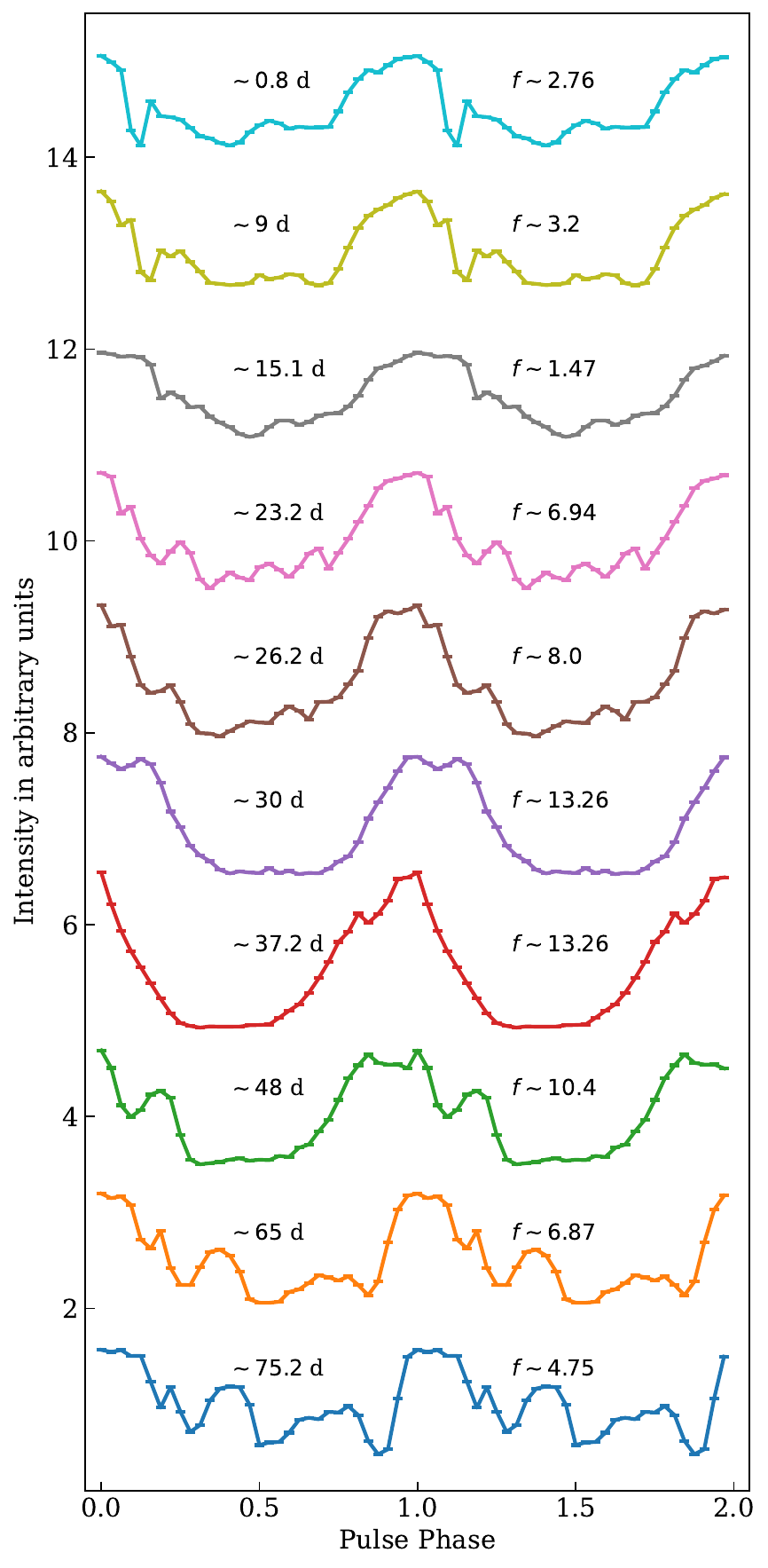}
\caption{The above figure presents the pulse profiles from different \nicer{} observations in 0.5--10 keV energy range. The time since MJD 59942 and the flux (in units of $10^{-9}$, \erg) of the corresponding observation has been marked with each pulse profile.}
	 \label{fig:profile}}
\end{figure}

The variation of pulsed fraction for different energy ranges is shown in Fig. \ref{fig:pulse-fraction} for different flux levels during the outburst. The pulsed fraction shows a trend to decrease with energy at different luminosity levels. The evolution of pulse profiles is shown in Fig. \ref{fig:profile} using \nicer{} observations, which indicates that the profiles are highly variable and luminosity dependent. The pulse profiles evolve from a multi-peak feature to a single-peak feature as flux increases. The profiles are closely sinusoidal and single peak at the peak of the outburst. In the decay phase of the outburst, the pulse profile evolves to a multi-peaked feature as luminosity decreases. The high variability of pulse profiles with luminosity indicates a change in beaming patterns which may be associated with the state transition of the source from subcritical to supercritical accretion regime.

%%%%%%%%%%%%%%%%%%%%%%%%%%%%%%%%%%%%%%%%%%%%%%%%%%

% Don't change these lines
\bsp	% typesetting comment
\label{lastpage}
\end{document}